# Gate-induced Superconductivity in atomically thin MoS$_2$ crystals.

Davide Costanzo[1], Sanghyun Jo[1], Helmuth Berger[2], and Alberto F. Morpurgo[1]

[1]DQMP and GAP, Université de Genève, 24 quai Ernest Ansermet, CH-1211 Geneva, Switzerland.

[2]Institut de Physique de la Matiere Complexe, Ecole Polytechnique Federale de Lausanne, CH-1015 Lausanne, Switzerland.

**When thinned down to the atomic scale, many layered van der Waals materials exhibit an interesting evolution of their electronic properties, whose main aspects can be accounted for by changes in the single-particle band structure. Phenomena driven by interactions are also observed, but identifying experimentally systematic trends in their thickness dependence is challenging. Here, we explore the evolution of gate-induced superconductivity in exfoliated MoS$_2$ multilayers ranging from bulk-like to individual monolayers. We observe a clear transition for all the thicknesses down to the ultimate atomic limit, providing the first demonstration of superconductivity in atomically thin exfoliated crystals. Additionally, we characterize the superconducting state by measuring the critical temperature ($T_C$) and magnetic field ($B_C$) in a large number of multilayer devices, upon decreasing their thickness. The superconducting properties change smoothly down to bilayers, and a pronounced reduction in $T_C$ and $B_C$ is found to occur when going from bilayers to monolayers, for which we discuss possible microscopic mechanisms. Finding that gate-induced superconductivity persists in individual monolayers, which form the basic building blocks of more sophisticated van der Waals heterostructures, opens new possibilities for the engineering of the electronic properties of materials at the atomic scale.**

The ability to produce few-atom-thick two-dimensional (2D) materials of excellent quality[1] [2] –such as graphene[3] [4] [5], semiconducting transition metal dichalcogenides (TMDs)[6] [7], and phosporene[8]– is an impressive breakthrough in condensed matter physics and nano-electronics. Upon adding an individual monolayer, the electronic properties of these 2D materials change drastically, so that multilayers of different thickness truly represent distinct electronic systems[9][10][11][12]. The basic aspects of their thickness-dependent properties are usually well-captured by single-particle band-structure calculations[13][14][15][16]. Depending on the specific system, experimental conditions, and physical processes investigated, however, interaction effects can also play an important role. Examples are excitonic effects[17][18][19] and gap renormalization in semiconducting TMDs[20], or the transport properties of few-layer, suspended graphene very close to charge neutrality[21][22][23]. As compared to properties described by a single-particle picture, exploring the thickness dependence of phenomena in which interactions play a key role is considerably more complex[24], and only limited work has been done.

Here, we report the investigation of gate-induced superconductivity in $MoS_2$ multilayers with a thickness ranging from six to one monolayer, and reveal a systematic behavior of the thickness dependence of the key properties characterizing the superconducting state, such as critical temperature ($T_C$) and field ($B_C$). Gate-induced superconductivity at the surface of semiconducting TMDs has been first demonstrated in recent pioneering work on $MoS_2$ field-effect transistors[25] (FETs) with liquid gates[26][27][28]. These experiments have been performed on thick exfoliated layers, behaving in all regards as bulk, and have led to the observation of $T_C$ values up to 12 K upon accumulation of electron surface densities close to $n \sim 10^{14}$ cm$^{-2}$. The relatively high critical temperature and the possibility to obtain chemically stable monolayers by simple exfoliation techniques make $MoS_2$ an ideal choice to investigate the evolution of superconductivity with reducing thickness down to the atomic scale.

Fig. 1a shows an optical microscope image of a typical device implemented on a six-layer (6L) $MoS_2$, ready to be covered with ionic liquid (see the scheme in Fig. 1b). Upon sweeping the gate voltage $V_G$ for positive values, electrons are accumulated, resulting in an increase of surface conductivity (Fig. 1c). At room temperature, measurements using different contact pairs as voltage probes give the same conductivity values (compare the three curves in Fig. 1c), which is indicative of a good device uniformity. This is the typical behavior of FETs based on $MoS_2$ multilayers of all thicknesses. In order to investigate the occurrence of superconductivity, we bias the ionic-liquid FET at $T \sim 220$K, by applying a gate voltage $V_G \sim$ 2-5 V (with the precise values depending on the specific device) before cooling it down slowly to $T = 1.5$ K. Fig. 1d compares the temperature dependence of the resistance below $T =$ 15 K measured on a 6L devices, with different contact pairs and in multiple cool-downs, at nominally comparable carrier density (1-3 × 10$^{14}$ cm$^{-2}$, see below). The expected superconducting transition is present in all curves and manifests itself in a sharp drop of the resistance (albeit not to a $R = 0$ state), which is shifted to lower temperature (and eventually entirely suppressed) by the application of a perpendicular magnetic field (Fig. 1e). The data show a rather broad range of values for the superconducting critical temperature, with the highest $T_C$ corresponding to the maximum values reported earlier[25] for devices realized on thick, bulk-like, $MoS_2$ flakes. The observed variations in the $R(T)$ curves and $T_C$ measured with different pairs of contacts are a manifestation of the carrier density non-uniformity that invariably appears at low temperature in these devices, very likely caused by the frozen ionic liquid locally detaching from the surface of $MoS_2$ (as we have recently discussed in more detail for $WS_2$ ionic-liquid gated FETs[29]). This non-uniformity makes comparing the behavior of multilayer of differ thickness complex, as it requires more devices to be investigated and multiple cool-downs to be performed, to extract representative data. Nevertheless, in the case of 6L $MoS_2$, measurements done with several different pairs of contacts exhibited a clear superconducting transition in all devices that were cooled down, whenever $n \sim 10^{14}$ cm$^{-2}$ or larger.

Having established the occurrence of robust gate-induced superconductivity in 6L $MoS_2$, with critical temperature and field comparable to those of bulk $MoS_2$, we discuss the behavior of thinner multilayers. Fig. 2a shows the low-temperature $R(T)$ curves measured at different magnetic field $B$ on a monolayer device exhibiting a clear superconducting transition to a $R=0$ $\Omega$ state; the response of the same device to a perpendicular magnetic field (at $T$=1.5 K) is shown in Fig. 2b. These observations demonstrate that gate-induced superconductivity in $MoS_2$ persists all the way down to the ultimate atomic scale. As compared to 6L $MoS_2$, the critical temperature in monolayers is significantly lower (the transition onset is just above 2 K) and the superconducting state is killed at much smaller $B$ values ($B_C \sim$ 0.05-0.1 T as compared to 5-10 T for the 6L device, see Fig. 1e; the critical field $B_C$ is defined as the magnetic field for which the measured resistance equals half of that in the normal state just above $T_C$). We conclude that –for approximately the same density of electrons and in the presence of comparable inhomogeneity (see Fig. 2c,d)– the superconducting state is weaker in monolayers than in 6L $MoS_2$. Indeed, whereas in 6L $MoS_2$ gate-induced superconductivity has been seen in every cool down and for all the working contact pairs, in monolayers observing the superconducting state is more difficult, and in several devices (cooled down multiple times) we observed no superconducting transition down to $T$=1.5 K (i.e., the lowest temperature reached in our experiments). These non-superconducting monolayers still exhibited metallic behavior (Fig. 2e), albeit with a higher resistivity than in the superconducting ones (for which the square resistance just above $T_C$ was approximately 100 Ω/square or less, rather than a few kilo-Ohms/square). Interestingly, the absence of superconductivity allows a clear weak-antilocalization signal due to phase-coherent single-electron transport to be observed upon the application of a magnetic field, consistently with the strong spin-orbit interaction present in $MoS_2$[30] (see Fig. 2f; a detailed analysis of this phenomenon will be discussed elsewhere).

For bilayers and thicker multilayers, the behavior of the superconducting transition resembles more closely the one found in bulk-like flakes, than that of monolayers. For instance, in bi and thicker layers we have found superconductivity in each device that we have cooled down, as long as $n \sim 10^{14}$ cm$^{-2}$ or larger. We illustrate this finding with data measured on bilayer $MoS_2$, shown in Fig. 3. The transition in this device is rather sharp with an onset close to $T_C = 7$ K. It is suppressed by applying a perpendicular magnetic field on a scale of a few Tesla (see Fig. 3a,b), which is comparable to what we found in 6L $MoS_2$ and much larger than the critical field found for monolayers. A very pronounced supercurrent is seen at 1.5 K, which is progressively suppressed by increasing $T$ or the applied perpendicular magnetic field (see Fig. 3c and its inset).

To explore the evolution of the superconducting state as a function of thickness in more detail, we compare the $R(T)$ and $R(B)$ curves measured in different multilayers, and search for systematic trends in the critical temperature and critical magnetic field. Representative $R(T)$ and $R(B)$ curves are plotted in Fig. 4a,b. In all cases we can identify the transition temperature $T_C$ and critical field $B_C$, even when the resistance at low $T$ does not vanish (which is a consequence of the large inhomogeneity in carrier density[29], as mentioned above). We find that the values of $T_C$ and $B_C$ in monolayers are significantly smaller than in thicker multilayers. This conclusion is confirmed by comparing data extracted from all the devices that we have measured (shown in Fig. 4c,d,e), which allow two main conclusions to be drawn. First, $T_C$ exhibits a slow decrease upon reducing the thickness from 6L to 2L, which appears to be statistically significant despite the spread in data due to carrier density inhomogeneity. Second, $T_C$ in monolayers is unambiguously suppressed as compared to all thicker multilayers, and a "jump" from approximately 7 K to 2 K is seen when passing from bilayer to monolayer $MoS_2$. This last conclusion is also apparent when plotting $T_C$ versus charge density (as inferred from Hall measurements): indeed, Fig. 4d shows that monolayers exhibit a significantly suppressed $T_C$ as compared to other multilayers, irrespective of carrier density and layer thickness. Finally, also when looking at the critical magnetic field, it is fully apparent that the values in monolayers are much smaller (more than one order of magnitude) than those extracted from all thicker multilayers (see Fig. 4e).

These results show that superconductivity in multilayer $MoS_2$ provides a clear example of an interaction effect exhibiting a systematic evolution as a function of thickness, all the way to the ultimate monolayer level. Finding that superconductivity is suppressed upon reducing the multilayer thickness may be expected: when the system becomes more two-dimensional, the (thermal and quantum) fluctuations responsible for the suppression of the long-range superconducting order become more important[31,32]. A complete understanding, however, requires the identification of specific microscopic mechanisms, and there are two candidates that likely play a major role. The first is associated to the physical thickness of the gate-induced accumulation layer at the surface of $MoS_2$. This thickness is determined by electrostatic screening, which is estimated to be approximately 1 nm[33]. A crossover can therefore be expected when multilayers thinner than the charge accumulation length are used. Since the bilayer thickness is about 1.5 nm[34], such a crossover indeed occurs when going from bilayer to monolayer. The second possible mechanism is related to the quantum mechanical states that are occupied by the electrostatically accumulated electrons. There is consensus, based on *ab-initio* calculations, that electrons added to the conduction band of $MoS_2$ monolayers initially fill states close to the K (and K') point[35,36,37], and that for thick multilayers electron accumulation first occurs at the Q-point[12,38,39], i.e. in a different part of the Brillouin zone. At what thickness the transition from K- to Q-point electron accumulation occurs, appears to depend on the approximations made in the calculations. Existing results strongly suggests that it takes place when passing from monolayer to bilayer[12,38,39,40], but the possibility that it occurs at larger thicknesses is not excluded[16,41,42] (taking properly into account the large perpendicular electric field unavoidably present at the surface in a transistor configuration is also important[42]). The transition would certainly affect the superconducting state, since the strength of the electron-phonon interaction and the density of states (which determine $T_C$[43]) around the K- and the Q-points are different, thus providing a realistic scenario to explain for the "weaker" superconductivity observed in $MoS_2$ monolayers.

It seems clear from these arguments that the occurrence of superconductivity in thick multilayers does not a priori implies that monolayers should also be superconducting. This conclusion reiterates the notion that $MoS_2$ multilayers of different thickness are in all regards distinct electronic systems, and underscore the relevance of investigating experimentally the thickness dependence of the superconducting transition. Irrespective of the specific mechanism responsible for the weakening of the superconducting state in $MoS_2$ monolayers, the gate-control of superconductivity in atomically thin layers demonstrated here is particularly interesting in the context of so-called van der Waals heterostructures[44]. Contrary to other interesting systems in which superconductivity was found to occur in vacuum-deposited, epitaxial films of atomic thickness (such as FeSe[45,46] and Pb[47,48]), $MoS_2$ is fully chemically stable in air, which makes it easy to use for the realization of artificial structures. As demonstrated by different recent experiments, the electronic properties of these structures can be engineered by suitably choosing the sequence of layers that are stacked together[44]. Including $MoS_2$ mono or bilayers in combination with ionic liquid gating adds superconductivity to the phenomena that can be investigated, controlled, or used to control the electronic properties of engineered van der Waals heterostructures.

## Methods

**Device fabrication.** Flakes of $MoS_2$ multilayers of different thickness were mechanically exfoliated from a bulk crystal using adhesive tape and transferred onto a $Si/SiO_2$ substrate. The thickness of the flakes was identified by measuring their optical contrast, in conjunction with atomic force microscopy and photoluminescence measurements. 50 nm Au electrodes were patterned by conventional nanofabrication techniques (e-beam lithography, metal evaporation, and lift-off) and annealed at 200 °C for two hours in a flow of a $Ar/H_2$ (100/10 sccm) mixture to reduce the contact resistance. Together with the electrodes, a large Au pad acting as a gate electrode for the ionic liquid was also deposited onto the substrate. PMMA resist was subsequently spun and patterned to open "windows" in correspondence of the $MoS_2$ flake, to define the region where the ionic liquid is in direct contact with the multilayer, i.e., the region where charge accumulation occurs upon the application of a voltage to the gate electrode. PMMA on top of the gate was also removed. The substrate was then mounted on a chip carrier and wire bonded, and a small droplet of ionic liquid DEME-TFSI (Kanto corporation) was placed onto the device in a glove box with controlled atmosphere (sub-ppm $O_2$ and $H_2O$ concentration). The device was rapidly transferred into the variable temperature insert (VTI) of a cryo-free Teslatron cryostat (Oxford instruments; base temperature of 1.5 K), where it was then left in vacuum ($10^{-6}$ mbar) for 1 day at room temperature to remove oxygen and humidity present in the ionic liquid, before starting the electrical measurements.

**Transport measurements.** All transport measurements discussed here were performed in vacuum or He atmosphere, inside the VTI of our Teslatron cryostat, equipped with a 12 T superconducting magnet. To investigate superconductivity upon accumulation of a high density of electrons, the device temperature was first set to 220 K, just above the freezing point of the ionic liquid, and a gate voltage was applied (typically between 2 and 5 V, depending on the measurement runs). As compared to room temperature, at $T = 220$ K possible chemical reactions between $MoS_2$ and the ionic liquid slow down, enabling a wider range of gate voltages to be applied without inducing device degradation. During gate biasing process, the gate leakage current was carefully monitored to ensure that it remained negligibly small at all times (below 1 nA). The device was then cooled down to the lowest base temperature of the system. The same devices have been cooled down and warmed up multiple times, changing the applied gate voltage at high temperature, without degradation. The transport characteristics were measured using a standard lock-in technique or in dc for *I-V* measurements (we used a Stanford SR830 lock-in amplifier, a Keithley 2400 source-meter, and an Agilent 34401a digital multimeter, in combination with homemade low-noise voltage and current sources, as well as current/voltage amplifiers).

**ACKNOWLEDGEMENTS**

We gratefully acknowledge A. Ferreira for technical help and T. Giamarchi, F. Mauri, and M. Calandra for useful discussions. Financial support from the Swiss National Science Foundation (SNF) and from the EU Graphene Flagship Project is gratefully acknowledged.

**AUTHOR CONTRIBUTIONS**

D.C. fabricated the majority of the devices and performed most of the measurements benefitting from the assistance and supervision of S.J.. D.C. and S.J. analyzed data. H.B. provided the $MoS_2$ crystals. A.F.M. proposed the experiment and supervised the research. All authors have discussed the results and contributed to their interpretation. D.C., S.J. and A.F.M. wrote the manuscript.

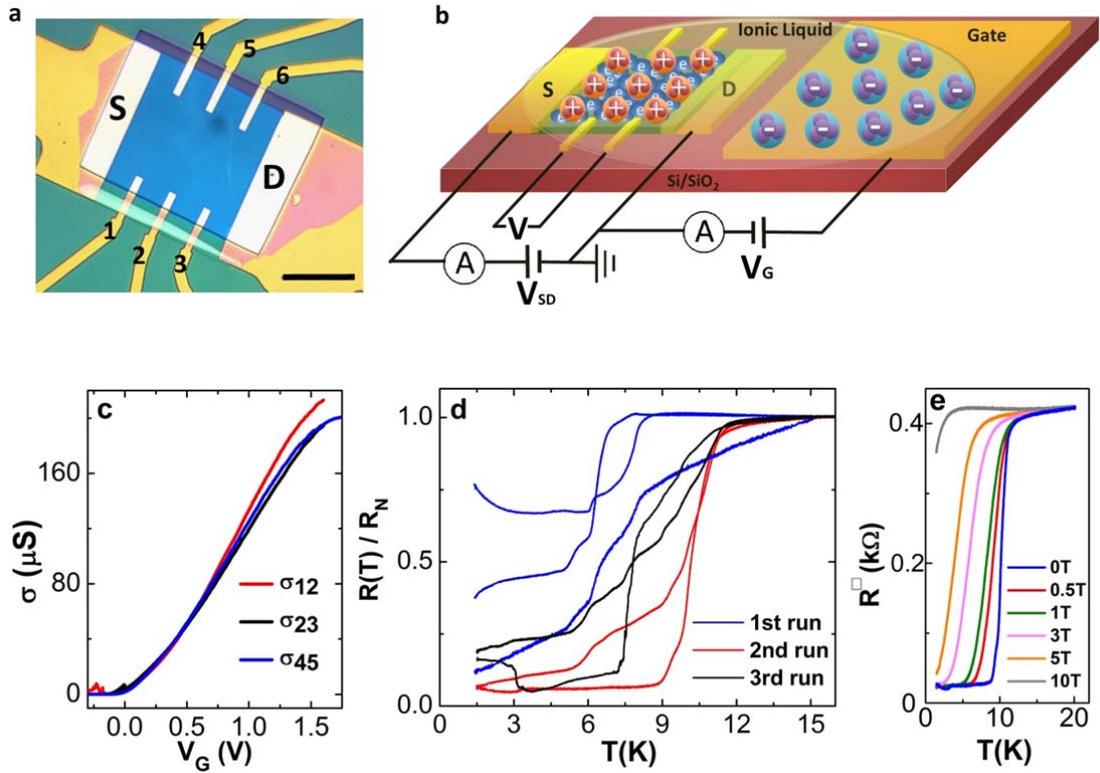

**Fig.1| Device characteristics and superconductivity in an ionic-liquid-gated six-layer MoS$_2$ transistor. a.** Optical microscope image of a 6L MoS$_2$ field effect transistor (FET), prior to deposition of the ionic liquid. The bar in the right-down corner is 15 μm long. In the experiments current is sent from the source (S) to the drain (D) electrode, while different voltage probes (labeled from 1 to 6) are used to measure the voltage. **b**. Schematic illustration of an ionic liquid (IL) gated FETs, under electron accumulation. **c**. Gate voltage ($V_G$) dependence of the room-temperature conductivity measured in different four-probe configurations ($\sigma_{AB}$ indicates the conductivity obtained from the voltage drop measured with probes A and B). The data shows that above the freezing temperature of the ionic liquid the device exhibits homogeneous behavior. **d**. Temperature dependence of the (normalized) four-probe resistance at high electron density ($n \sim 10^{14}$ cm$^{-2}$) with clear manifestations of the superconducting transition. The different colors correspond to data taken in different cool downs (in each cool down different curves are measured with different contact pairs). The broad transitions and the $T_C$ variations observed when measuring with different contact pairs are due to the inhomogeneity arising from the local detachment of the frozen ionic liquid. **e**. A progressive suppression of the transition is observed upon increasing the perpendicular magnetic field, substantiating the occurrence of superconductivity (the critical field in this device is $B_C \sim 7$ T).

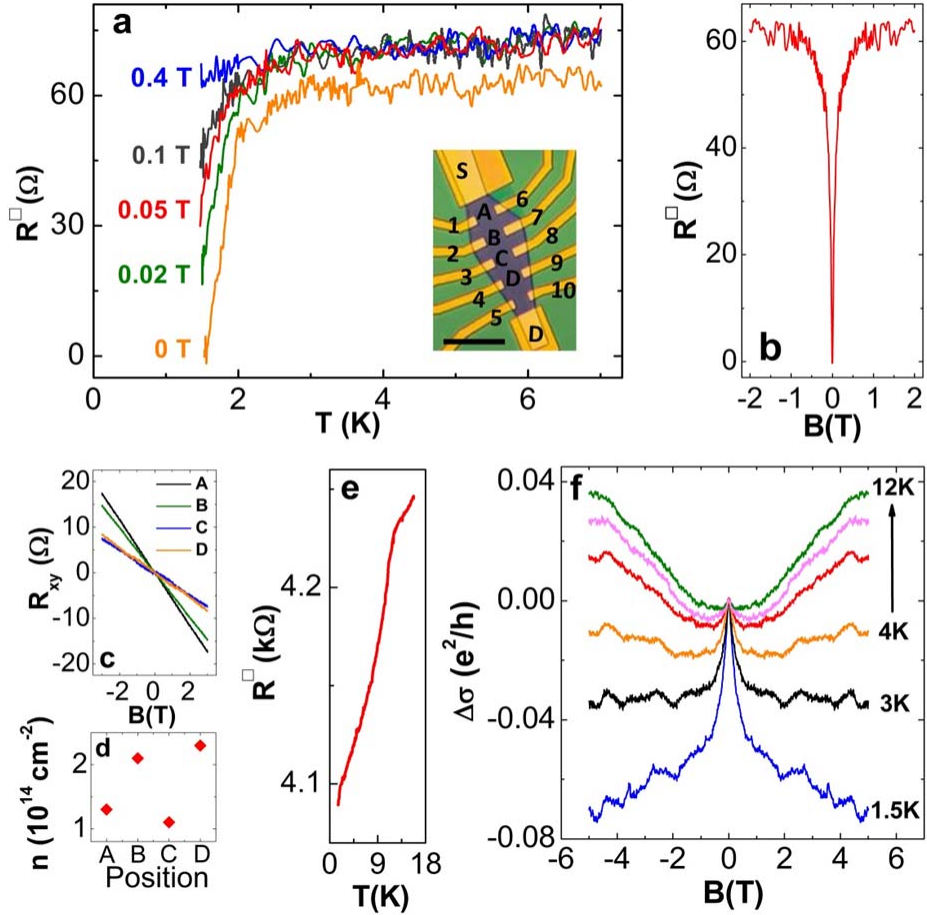

**Fig.2| Gate-induced superconductivity in monolayer MoS$_2$. a.** Temperature dependence of four-probe square resistance ($R^\square$) at high electron density ($n \sim 10^{14}$ cm$^{-2}$) in the monolayer device shown in the inset (the black bar in the corner is 15 µm long), for different values of perpendicular magnetic field. The sharp drop to zero of the resistance below $T \simeq 2$ K at $B = 0$ T demonstrates the occurrence of superconductivity. Despite the accumulation of a high electron density ($n \sim 10^{14}$ cm$^{-2}$), superconductivity is observed only with one pair of contacts (7 - 8), contrary to the case of the 6L device, in which a superconducting transition is observed irrespective of the contacts used. The superconducting transition is suppressed in a relatively small magnetic field, consistently with the magnetoresistance data shown in **b** (critical field $B_C \sim 0.1$ T). **c**. Hall effect measurements and the corresponding carrier density (**d**) measured with different pair of contacts at positions A, B, C, and D (defined in the inset of **a**), showing fluctuations as large as a factor of 2-3. **e**. For the contact pairs in which superconductivity is not observed, a metallic behavior of the resistance upon lowering temperature is visible, and the magnetoconductivity exhibits a clear weak-antilocalization behavior, as shown in (**f**).

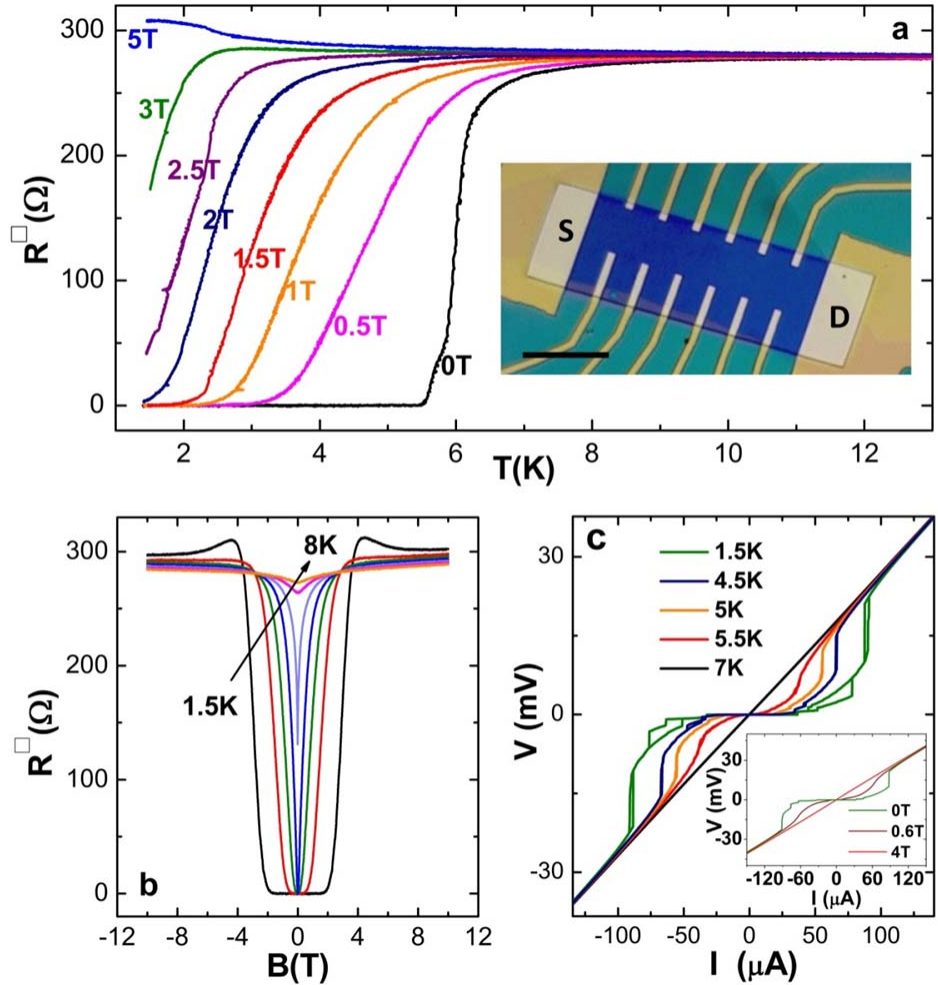

**Fig.3| Superconductivity in bilayer MoS$_2$. a.** Temperature dependence of the four-probe square resistance measured at an applied gate voltage $V_G$ = 2.2 V, for different values of perpendicular magnetic field $B$. At $B$ = 0 T, the superconducting transition occurs at $T_C \approx 7$ K, much higher than in the single layer case, and at $T$=1.5 K, the zero resistance state remains visible up to $B \approx 2$ T (see also the magnetoresistance data in **b**), again significantly higher than in the monolayer. In the inset, an optical image of the device is shown (the scale bar corresponds to 15 μm) **c.** *I-V* characteristic showing a pronounced supercurrent at $T$ = 1.5 K, which is gradually suppressed upon increasing the temperature (or the magnetic field, as shown in the inset).

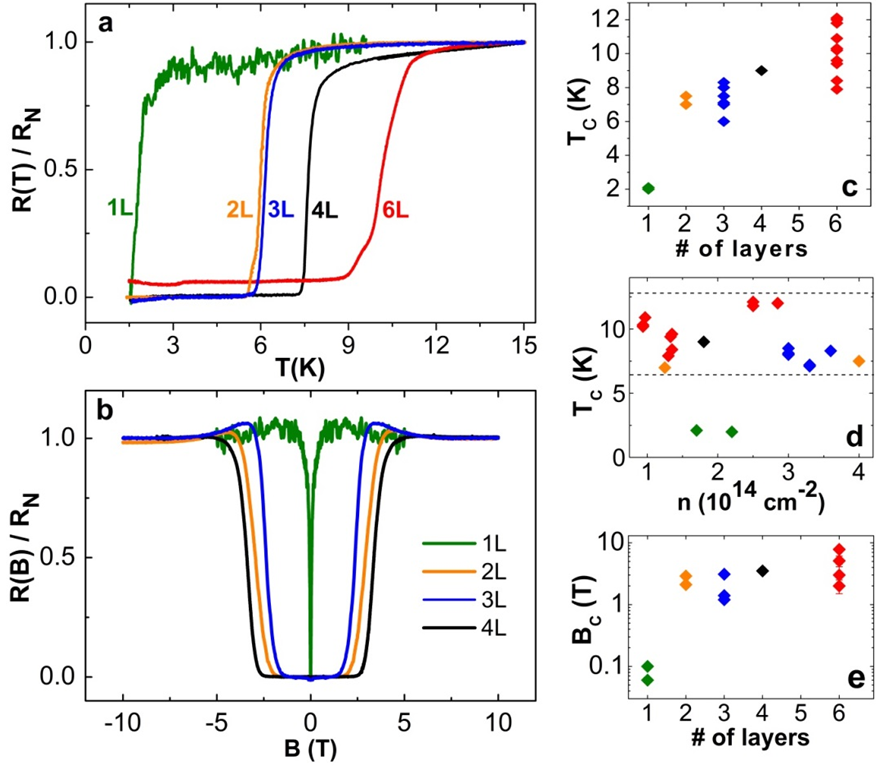

**Fig. 4: Evolution of superconductivity in MoS$_2$ upon decreasing thickness down to the atomic scale. a.** Comparison of superconducting transitions in devices of different thickness, from 1L to 6L, as a function of temperature and (in **b**) of magnetic field (at T=1.5 K). **c.** Summary of the critical temperature values in all MoS$_2$ devices of different thickness (from 1ML to 6ML) that we have measured at electron density $n \sim 1 \times 10^{14}$ cm$^{-2}$ or higher. Despite the device-to-device fluctuations, a trend is visible: $T_C$ decreases gradually as the thickness is reduced from 6L to 2L, and exhibits a "jump" –from approximately 7 K to 2 K– when passing from bilayers to monolayers (in **c-e** the color of each symbol indicates the multilayer thickness, according to the same color code used in **a**). **d.** The value of $T_C$ for all observed superconducting transitions is plotted versus the respective carrier density. No clear relation between $T_C$ and $n$ is seen in the data, because of the carrier density inhomogeneity (see discussion in the main text). The data nevertheless show that, irrespective of $n$, $T_C$ in monolayers is significantly lower than for thicker multilayers. **e.** The behavior of the critical field (plotted in log scale) as a function of thickness parallels that of $T_C$: a slow decrease is seen upon decreasing thickness from 6L to 2L, followed by a one-order-of-magnitude reduction in passing from bilayer to monolayers.

# Supplementary Information
# Gate-induced Superconductivity in atomically thin MoS$_2$ crystals.

Davide Costanzo[1], Sanghyun Jo[1], Helmuth Berger[2], and Alberto F. Morpurgo[1]

[1]*DQMP and GAP, Université de Genève, 24 quai Ernest Ansermet, CH-1211 Geneva, Switzerland.* [2]*Institut de Physique de la Matiere Complexe, Ecole Polytechnique Federale de Lausanne, CH-1015 Lausanne, Switzerland.*

**Analysis of the *I-V* curves of superconducting 4L MoS$_2$ in terms of a Berezinskiĭ-Kosterlitz-Thouless scenario.**

The electrostatically induced superconducting state at the surface of an ionic-liquid gated MoS$_2$ transistor is expected to have a two-dimensional (2D) character, since the electrostatic screening length at a carrier density of $10^{14}$ cm$^{-2}$ is approximately 1 nm[1]. According to the Mermin-Wagner theorem, in 2D systems, long-range order is suppressed by thermal fluctuations[2]; quasi-long-range ordering, however, can still occur. This quasi long-range ordering is characteristic of the Berezinskiĭ-Kosterlitz-Thouless (BKT) mechanism for superconductivity, through which a coherent superconducting state occurs at a temperature $T_{BKT}$, lower than the mean-field critical temperature[3][4].

There are different experimental manifestations of the BKT regime that can be looked for in the experiments, to see whether the observed behavior is compatible with that expected for 2D superconductivity. A well-known aspect of the BKT transition is the occurrence of power-law *I-V* characteristics: $V \propto I^\alpha$ where $\alpha = 1$ for *T* well above $T_{BKT}$, $\alpha = 3$ for $T = T_{BKT}$, and $\alpha > 3$ for $T < T_{BKT}$ [5][6]. We have looked for such a behavior in the *I-V* curves measured in our devices. Fig. S1a shows the *I-V* characteristic of a 4L MoS$_2$ FET (in log-log scale) measured below the onset temperature of superconductivity. A linear regime is clearly visible, whose slope $\alpha$ (the exponent of the power law) increases upon lowering temperature, as expected. The exponent $\alpha$ is obtained from fits of the slope for each temperature (black lines in Fig. S1a), and plotted as a function of temperature in Fig. S1b. $\alpha$ increases with lowering *T* for temperatures well below the onset temperature, and $\alpha = 3$ at $T = 7.4$ K, providing a first estimate of $T_{BKT}$.

To confirm the BKT nature of the transition, we also looked at the temperature dependence of the resistance close to $T_{BKT}$. Theory predicts that $R = R_0 \exp(-bt^{-1/2})$, in contrast to the power law dependence expected in the Ginzburg-Landau theory ($R_0$ and $b$ are material dependent parameters, and $t = T/T_{BKT} - 1)^{5,6}$. This functional dependence can be tested by checking that $[d(\ln R)/dT]^{-2/3}$ varies linearly with $T$ above $T_{BKT}$, and that it extrapolates at 0 for $T = T_{BKT}$. The data, shown in Fig. S1c, indeed displays the expected behavior. By extrapolating to zero the linear part of $[d(\ln R)/dT]^{-2/3}$ we find that $T_{BKT} = 7.4$ K, in good agreement with the previous estimation from the $I$-$V$ characteristic. The analysis of our experimental data is thus compatible with the theoretical expectation for a 2D BKT superconducting transition.

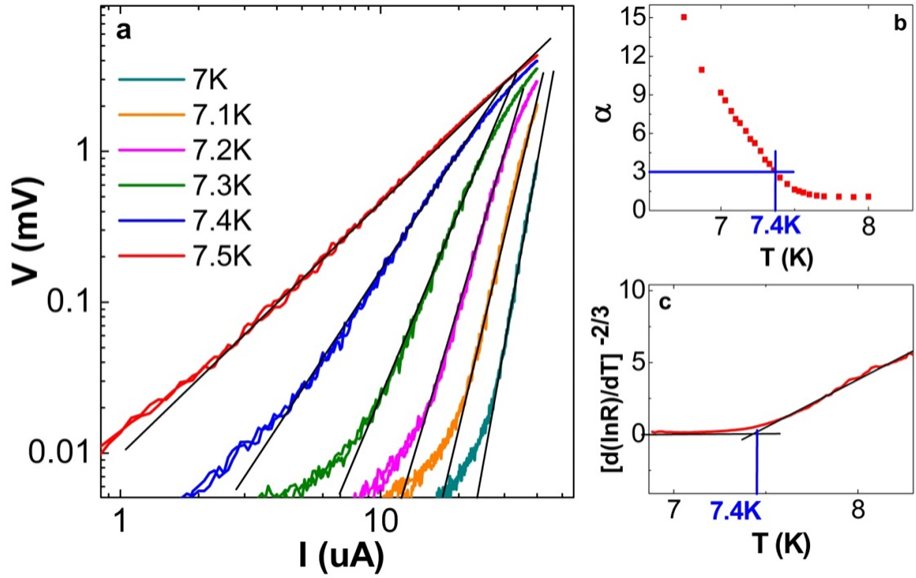

**Fig. S1. Berenzinskiĭ-Kosterlitz-Thouless behavior in an ion-liquid gated 4L MoS$_2$ FET. a.** Power-law dependence $V \propto I^\alpha$ of the $I$-$V$ characteristics for different temperatures (below the onset of superconductivity) plotted in log-log scale. The solid black lines are best fits done to extract the exponent $\alpha$. **b.** Temperature dependence of $\alpha$. $T_{BKT}$ corresponds to the value of $T$ for which $\alpha = 3$, leading to $T_{BKT} = 7.4$ K. **c.** Linear behavior of $[d(\ln R)/dT]^{-2/3}$ as a function of temperature. The extrapolation to zero (black solid line) provides an independent estimate of $T_{BKT}$, which also gives 7.4 K, consistently with the analysis of $I$-$V$ characteristics.